\documentclass[12pt]{article}

\usepackage{sbc-template}
\usepackage{multirow}
\usepackage{graphicx,url}
\usepackage{tabularx}
\usepackage{float}
\usepackage[utf8]{inputenc}
\usepackage{xcolor}
\usepackage{tikz}
\usepackage{wrapfig}
\usetikzlibrary{fit,backgrounds,positioning,arrows.meta}
\definecolor{humano}{HTML}{C2622E}
\definecolor{ia}{HTML}{3F6FB0}

\hypersetup{hidelinks}

\title{What Does It Take to Research with AI? A Rapid Review of Competencies to Train LLM-Literate Researchers}

\author{Danilo Monteiro Ribeiro\inst{1}, Ronnie de Souza Santos\inst{1,4},
  Rodrigo Siqueira\inst{1},\\
  Breno Andrade\inst{1}, Rafael Batista Duarte\inst{1,2}, Gilberto Hida\inst{1},\\
  Julia Alencar\inst{1}, Gustavo Pinto\inst{3}}

\address{CESAR School -- Recife, PE -- Brazil
\nextinstitute
  Universidade de Pernambuco (UPE) -- Recife, PE -- Brazil
\nextinstitute
  Federal University of Pará (UFPA) -- Belém, PA -- Brazil
\nextinstitute
  University of Calgary -- Calgary, AB -- Canada
}

\begin{document}

\maketitle

\begin{abstract}
The growing adoption of Large Language Models in scientific research has created a need to understand what competencies researchers and graduate students require to use these tools critically and responsibly. This rapid review analyzed 194 articles retrieved from Elicit and Google Scholar (2022--2025), from which 40 were selected for competency extraction and thematic analysis following independent dual screening (Gwet AC1: 0.76--0.83). Eight competencies were identified. The most prevalent was domain expertise and oversight of AI outputs ($\Sigma n = 123$), encompassing subject-matter mastery, systematic skepticism, source verification, and researcher accountability. Other key competencies include metacognition and decision making about AI use ($\Sigma n = 55$), ethics and academic integrity ($\Sigma n = 53$), prompt engineering for research ($\Sigma n = 38$), and reproducibility of AI use ($\Sigma n = 29$). AI literacy and technical knowledge ($\Sigma n = 16$) was explicitly identified as a risk factor when absent, with domain expertise treated as a prerequisite for meaningful critical evaluation. The findings suggest that preparing researchers to use LLMs goes beyond technical instruction, requiring an integrated set of epistemic, ethical, and methodological competencies centered on human accountability for the knowledge produced. These results have direct implications for the design of graduate programs and AI literacy initiatives.
\end{abstract}

\begin{resumo}
A crescente adoção de Large Language Models na pesquisa científica criou a necessidade de compreender quais competências pesquisadores e estudantes de pós-graduação precisam desenvolver para utilizar essas ferramentas de forma crítica e responsável. Esta rapid review analisou 194 artigos recuperados do Elicit e do Google Scholar (2022--2025), dos quais 40 foram selecionados para extração de competências e análise temática, após triagem dupla independente (Gwet AC1: 0,76--0,83). Oito competências foram identificadas. A mais prevalente foi o domínio do assunto e a supervisão das saídas da IA ($\Sigma n = 123$), que abrange conhecimento do tema, ceticismo sistemático, verificação de fontes e responsabilidade do pesquisador sobre o conteúdo produzido. Outras competências relevantes incluem metacognição e tomada de decisão sobre o uso da IA ($\Sigma n = 55$), ética e integridade acadêmica ($\Sigma n = 53$), engenharia de prompt para pesquisa ($\Sigma n = 38$) e reprodutibilidade do uso da IA ($\Sigma n = 29$). A alfabetização em IA e o conhecimento técnico ($\Sigma n = 16$) foram explicitamente identificados como fator de risco quando ausentes, com o conhecimento de domínio atuando como pré-requisito para uma avaliação crítica significativa. Os resultados sugerem que preparar pesquisadores para o uso de LLMs vai além da instrução técnica, exigindo um conjunto integrado de competências epistêmicas, éticas e metodológicas centradas na responsabilidade humana pelo conhecimento produzido. Esses resultados têm implicações diretas para o design de programas de pós-graduação e iniciativas de alfabetização em IA.
\end{resumo}

\section{Introduction}

Artificial Intelligence (AI) has supported teaching, learning, and assessment for years \cite{chen2020artificial}, stimulating discussions on digital literacy, AI literacy, critical thinking, and the skills needed in technology-mediated environments \cite{ng2021conceptualizing, reyes2021research, meerah2012measuring}. Generative AI and Large Language Models (LLMs) extended this through natural language interfaces \cite{meyer2023chatgpt, rahman2023chatgpt}. Unlike earlier educational technologies built for specific purposes, LLMs support a broad range of academic activities, from information retrieval and summarization to writing, programming, and problem solving \cite{meyer2023chatgpt, rahman2023chatgpt, lissack2024navigating}.

Software engineering education is one domain where this impact has drawn particular attention, with reported use in programming, requirements, design, testing, documentation, and learning activities \cite{kozov2024practical, kirova2024software}. Concerns have followed regarding academic integrity, overreliance on generated content, learning outcomes, and the preparation of professionals able to work responsibly in AI-assisted environments \cite{santos2026llm, kirova2024software, kozov2024practical}.

Beyond education, LLMs are increasingly used across the research lifecycle for literature exploration, idea generation, academic writing, coding, peer review, and data analysis \cite{rahman2023chatgpt, meyer2023chatgpt, reddy2025towards, trinkenreich2025get, trinkenreich2026taking}. This creates opportunities but also raises concerns regarding transparency, reproducibility, hallucinations, bias, source verification, scientific integrity, and researcher accountability \cite{lissack2024navigating, blau2024protecting, meyer2023chatgpt, rahman2023chatgpt}. In software engineering research, guidelines have begun to support the rigorous use and reporting of LLM-based studies \cite{baltes2025guidelines, trinkenreich2025get, trinkenreich2026taking}.

Research traditionally requires competencies related to critical evaluation of evidence, methodological rigor, information management, ethical conduct, knowledge synthesis, and scientific communication \cite{meerah2012measuring, reyes2021research, wohlin2007empirical, pizard2021training, kitchenham2004evidence, kitchenham2009systematic}, while AI literacy research emphasizes understanding AI capabilities and limitations, evaluating outputs, and making informed decisions about their use \cite{ng2021conceptualizing}. How these competencies intersect in AI-assisted research remains underexplored, and those required for the effective, critical, and responsible use of LLMs remain dispersed across the literature \cite{meyer2023chatgpt, rahman2023chatgpt, lissack2024navigating, trinkenreich2026taking}.

To address this gap, this study identifies and consolidates the competencies required for the use of LLMs in scientific research, with particular attention to software engineering research, analyzing studies published between 2022 and 2025 that document real practices mediated by generative AI tools. It addresses the following research question: \textbf{RQ: What competencies are required for the effective, critical, and responsible use of large language models throughout the scientific research process?} The paper makes three contributions: it synthesizes empirical evidence on how LLMs are used across the research process, identifies and consolidates competencies associated with AI-assisted research, and discusses implications for software engineering research, graduate education, and researcher preparation.

\section{Background}

Scientific research is fundamentally a process of producing, evaluating, interpreting, and communicating knowledge. In software engineering, evidence-based approaches emphasize systematic, transparent, and reproducible procedures \cite{kitchenham2004evidence, kitchenham2009systematic, wohlin2007empirical}, which are especially important in secondary studies such as systematic literature reviews, where conclusions depend on rigorous identification, selection, analysis, and synthesis of evidence \cite{kitchenham2004evidence, kitchenham2009systematic, pizard2021training}. Research quality therefore depends not only on the evidence but also on the decisions researchers make throughout the process.

LLMs have introduced new possibilities for supporting research, with reported use for literature exploration, idea generation, scientific discovery, coding, academic writing, peer review, and data analysis \cite{rahman2023chatgpt, meyer2023chatgpt, reddy2025towards, trinkenreich2025get, trinkenreich2026taking}. In software engineering research, these tools have been incorporated into exploratory investigations, empirical studies, and evidence synthesis \cite{trinkenreich2025get, trinkenreich2026taking}, helping researchers navigate increasingly large bodies of literature \cite{rahman2023chatgpt, meyer2023chatgpt, reddy2025towards}.

At the same time, this growing role has prompted discussions about researcher oversight, accountability, and the evaluation of outputs whose internal decision processes are difficult to inspect \cite{lissack2024navigating, blau2024protecting}. Reported concerns include hallucinations, inaccurate references, unsupported claims, bias, and limited transparency \cite{meyer2023chatgpt, rahman2023chatgpt, lissack2024navigating}. Such issues are particularly relevant in evidence synthesis, where judgments made during study selection, quality assessment, data extraction, coding, and synthesis directly shape the resulting body of evidence, and errors may propagate to subsequent analysis and interpretation \cite{kitchenham2004evidence, kitchenham2009systematic, baltes2025guidelines}.

Recent guidance for LLM-based software engineering studies reflects these concerns by emphasizing transparent reporting of AI use, reproducibility, documentation of prompts and interactions, and careful evaluation of outputs \cite{baltes2025guidelines}. Rather than treating LLMs solely as productivity tools, the literature increasingly examines their influence on retrieval, interpretation, synthesis, and communication \cite{lissack2024navigating, blau2024protecting, trinkenreich2026taking}. Consequently, AI-assisted research must preserve scientific rigor, maintain accountability for research decisions, and keep conclusions grounded in evidence.

\section{Method}

This study aims to identify and consolidate the competencies required for the use of LLMs in scientific research, with particular attention to software engineering research. To achieve this objective, we conducted a rapid review, an evidence synthesis approach that follows the main principles of systematic reviews while enabling timely investigation of emerging topics \cite{cartaxo2020rapid}. The review comprised four phases: study identification, study selection, data extraction, and thematic analysis.

\subsection{Study Selection}

Initially, 194 articles were selected for analysis, of which 100 were obtained from the first 10 pages of Google Scholar results and 94 were retrieved using the Elicit tool, already excluding duplicates also returned by Google Scholar. In Elicit, an AI-assisted semantic search was conducted in scientific databases, with studies ranked according to their thematic relevance. In Google Scholar, the first 10 pages of results were considered to broaden the coverage of the literature, including possible grey literature, preprints, and relevant publications not indexed in traditional databases. We used Elicit and Google Scholar rather than curated databases such as Scopus, Web of Science, the ACM Digital Library, and IEEE Xplore because, as a rapid review of a fast-moving topic, we prioritized timely retrieval over exhaustive coverage. The complete search strings and the protocol adopted in this stage are reported in the artifacts (Section~\ref{sec:artifacts}).

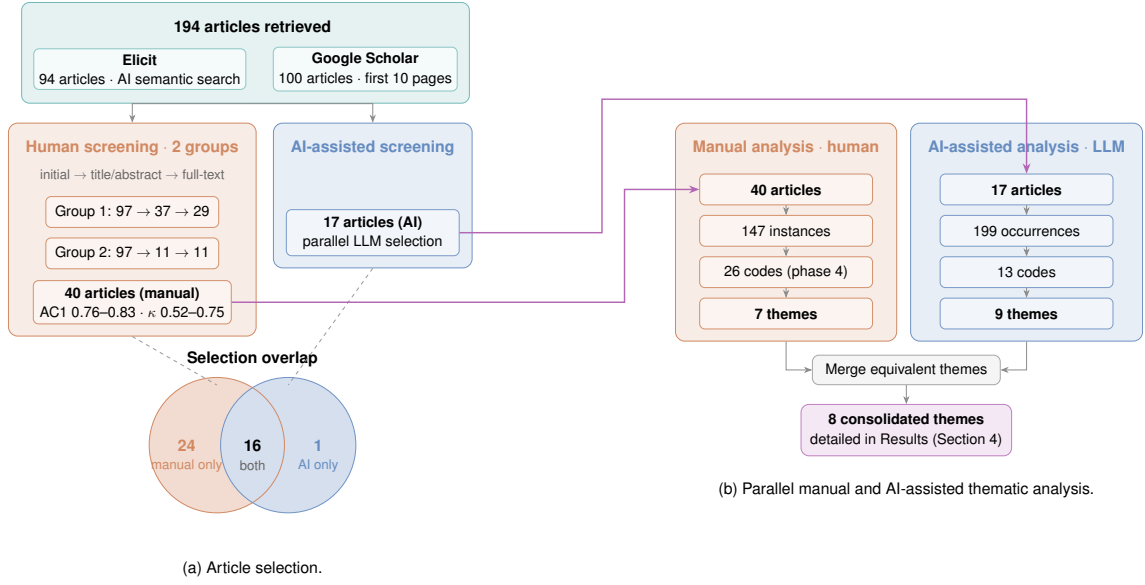
\begin{figure*}[!t]
\centering
\resizebox{\textwidth}{!}{%
\begin{tikzpicture}[
  font=\sffamily\small, >=Stealth,
  sub/.style={rounded corners=4pt, draw=teal!45, fill=teal!4, align=center, inner sep=4pt, minimum width=4.6cm},
  retr/.style={rounded corners=6pt, draw=teal!55, fill=teal!10, line width=0.7pt},
  hbox/.style={rounded corners=4pt, draw=humano!70, fill=humano!7, align=center, inner sep=5pt, minimum width=4.6cm, minimum height=0.82cm},
  abox/.style={rounded corners=4pt, draw=ia!70, fill=ia!7, align=center, inner sep=5pt, minimum width=4.6cm, minimum height=0.82cm},
  hcont/.style={rounded corners=7pt, draw=humano!75, fill=humano!12, line width=0.7pt},
  acont/.style={rounded corners=7pt, draw=ia!75, fill=ia!12, line width=0.7pt},
  fin/.style={rounded corners=6pt, draw=violet!50, fill=violet!9, align=center, inner sep=7pt, minimum width=5.4cm},
  mrg/.style={rounded corners=5pt, draw=black!35, fill=black!4, align=center, inner sep=6pt, minimum width=5.0cm},
  ttl/.style={font=\sffamily\bfseries},
  arr/.style={->, draw=black!45, line width=0.7pt},
  dashln/.style={dashed, draw=black!40, line width=0.5pt},
  flow/.style={->, draw=violet!60, line width=1.1pt},
  capt/.style={font=\sffamily\footnotesize, text=black!60, align=center},
  lab/.style={font=\sffamily},
]

\begin{scope}[shift={(-8.7,0)}]
  \node[sub] (elicit) at (-3.0,0) {\textbf{Elicit}\\[1pt]94 articles $\cdot$ AI semantic search};
  \node[sub] (gscholar) at (3.0,0) {\textbf{Google Scholar}\\[1pt]100 articles $\cdot$ first 10 pages};
  \node[ttl] (title) at (0,1.15) {194 articles retrieved};
  \begin{scope}[on background layer]
    \node[retr, fit=(title)(elicit)(gscholar), inner sep=9pt] (corpus) {};
  \end{scope}
  \node[ttl, text=humano!75] (th) at (-3.2,-2.1) {Human screening $\cdot$ 2 groups};
  \node[capt] (ths) at (-3.2,-2.8) {initial $\to$ title/abstract $\to$ full-text};
  \node[hbox] (g1) at (-3.2,-3.8) {Group 1: 97 $\to$ 37 $\to$ 29};
  \node[hbox] (g2) at (-3.2,-4.9) {Group 2: 97 $\to$ 11 $\to$ 11};
  \node[hbox] (hsum) at (-3.2,-6.2) {\textbf{40 articles (manual)}\\[1pt]AC1 0.76--0.83 $\cdot$ $\kappa$ 0.52--0.75};
  \begin{scope}[on background layer]
    \node[hcont, fit=(th)(ths)(g1)(g2)(hsum), inner sep=9pt] (hc) {};
  \end{scope}
  \node[ttl, text=ia!80] (ta) at (3.2,-2.1) {AI-assisted screening};
  \node[abox] (aibox) at (3.2,-4.35) {\textbf{17 articles (AI)}\\[1pt]parallel LLM selection};
  \begin{scope}[on background layer]
    \node[acont, fit=(ta)(aibox), inner sep=9pt] (ac) {};
  \end{scope}
  \draw[arr] (corpus.south) -| (hc.north);
  \draw[arr] (corpus.south) -| (ac.north);
  \node[ttl] (vt) at (0,-7.7) {Selection overlap};
  \begin{scope}[shift={(0,-10.0)}]
    \fill[humano!22] (-0.95,0) circle (1.8);
    \fill[ia!22] (0.95,0) circle (1.8);
    \draw[humano!75, line width=0.7pt] (-0.95,0) circle (1.8);
    \draw[ia!75, line width=0.7pt] (0.95,0) circle (1.8);
    \node[font=\sffamily\bfseries, text=humano!75] at (-1.75,0) {24};
    \node[font=\sffamily\bfseries] at (0,0) {16};
    \node[font=\sffamily\bfseries, text=ia!80] at (1.75,0) {1};
    \node[capt, text=humano!75] at (-1.75,-0.55) {manual only};
    \node[capt] at (0,-0.55) {both};
    \node[capt, text=ia!80] at (1.75,-0.55) {AI only};
  \end{scope}
  \draw[dashln] (hc.south) -- (-0.95,-8.4);
  \draw[dashln] (ac.south) -- (0.95,-8.4);
  \node[lab] at (0,-13.3) {(a) Article selection.};
\end{scope}

\begin{scope}[shift={(8.7,-2.6)}]
  \node[ttl, text=humano!75] (tm) at (-3.2,0.5) {Manual analysis $\cdot$ human};
  \node[hbox] (m0) at (-3.2,-0.6) {\textbf{40 articles}};
  \node[hbox] (m1) at (-3.2,-1.7) {147 instances};
  \node[hbox] (m2) at (-3.2,-2.8) {26 codes (phase 4)};
  \node[hbox] (m3) at (-3.2,-3.9) {\textbf{7 themes}};
  \begin{scope}[on background layer]
    \node[hcont, fit=(tm)(m0)(m1)(m2)(m3), inner sep=9pt] (mc) {};
  \end{scope}
  \node[ttl, text=ia!80] (tb) at (3.2,0.5) {AI-assisted analysis $\cdot$ LLM};
  \node[abox] (a0) at (3.2,-0.6) {\textbf{17 articles}};
  \node[abox] (a1) at (3.2,-1.7) {199 occurrences};
  \node[abox] (a2) at (3.2,-2.8) {13 codes};
  \node[abox] (a3) at (3.2,-3.9) {\textbf{9 themes}};
  \begin{scope}[on background layer]
    \node[acont, fit=(tb)(a0)(a1)(a2)(a3), inner sep=9pt] (acb) {};
  \end{scope}
  \draw[arr] (m0)--(m1); \draw[arr] (m1)--(m2); \draw[arr] (m2)--(m3);
  \draw[arr] (a0)--(a1); \draw[arr] (a1)--(a2); \draw[arr] (a2)--(a3);
  \node[mrg] (merge) at (0,-5.4) {Merge equivalent themes};
  \draw[arr] (mc.south) |- (merge.west);
  \draw[arr] (acb.south) |- (merge.east);
  \node[fin] (cons) at (0,-7.0) {\textbf{8 consolidated themes}\\[1pt]detailed in Results (Section 4)};
  \draw[arr] (merge) -- (cons);
  \node[lab] at (0,-8.6) {(b) Parallel manual and AI-assisted thematic analysis.};
\end{scope}

\draw[flow] (hsum.east) -- (1.2,-6.2) -- (1.2,-3.2) -- (m0.west);
\draw[flow] (aibox.east) -- (0.6,-4.35) -- (0.6,-0.8) -- (11.9,-0.8) -- (a0.north);

\end{tikzpicture}%
}
\caption{Overview of the review process}
\label{fig:methodological_execution}
\label{fig:analysis_pipeline}

{\footnotesize \textit{Note.} (a) Article selection: 194 retrieved articles were screened independently by humans (40 selected) and by an AI-assisted procedure (17 selected). (b) Thematic analysis: 7 manual and 9 AI-assisted themes were merged into 8 consolidated competencies.\par}
\end{figure*}

After the initial selection of articles from the respective search sources, the research team, composed of four researchers, met to define the screening strategy. The screening process was organized into two distinct groups, each composed of two researchers. Group 1 was responsible for analyzing the first 97 articles, while Group 2 was responsible for analyzing the remaining 97 articles. The methodological execution of the selection and screening process is summarized in Figure~\ref{fig:methodological_execution}(a).

The screening of studies was conducted based on previously defined inclusion and exclusion criteria. Studies were included if they: (I1) were published between 2022 and 2025 and documented scientific research practices mediated by generative AI tools; (I2) reported at least one concrete use of AI within a stage of the research process; (I3) described AI tools together with their purpose, research context, or level of autonomy; (I4) provided sufficient information to infer methodological adaptations or competencies associated with AI use; and (I5) were written in English or Portuguese, the paper that not attend the inclusion criteria are excludes.

In the first screening stage, the researchers assessed the titles and abstracts of the selected studies to determine their relevance to the use of artificial intelligence in scientific research. This stage resulted in 48 articles selected for full-text analysis, 37 from Group 1 and 11 from Group 2. In the second stage, these articles were analyzed in greater depth according to the inclusion and exclusion criteria defined for the study. As a result, 40 articles were selected for data extraction.

To complement the manual screening, an AI-supported validation was conducted on a set of 194 articles using the same inclusion and exclusion criteria. The application identified 17 articles already included in the manual review and 1 additional article not previously selected by the researchers. However, after reviewing this article, the researchers decided not to include it in the final corpus, as it did not fully meet the acceptance criteria. Therefore, the final set of studies remained composed of 40 articles.

To evaluate the reliability of the screening process, three chance-corrected agreement coefficients were computed over the reviewers' independent include/exclude classifications: Gwet's AC1 \cite{gwet2008computing}, Krippendorff's Alpha \cite{krippendorff2018content}, and Weighted Cohen's Kappa \cite{cohen1968weighted}. Gwet's AC1 values ranged from 0.76 to 0.83, indicating substantial to almost perfect agreement between reviewers, following the interpretation benchmarks proposed by Landis and Koch \cite{landis1977measurement}. Krippendorff's Alpha and Weighted Cohen's Kappa values ranged from 0.52 to 0.75, indicating moderate to substantial agreement. In this study, Gwet's AC1 was adopted as the primary agreement indicator because it is less sensitive to prevalence effects and category imbalance than traditional kappa-based statistics \cite{gwet2008computing}.

\subsection{Data Extraction and Data Analysis}

The extracted competency instances were analyzed using thematic analysis \cite{cruzes2011recommended}. We adopted an inductive approach in which themes emerged from the extracted data. Coding of the 147 competency instances followed an iterative process of comparison, refinement, and merging of codes representing similar skills, knowledge areas, responsibilities, or practices, which produced 26 codes in the fourth coding phase. These codes were subsequently consolidated into 7 higher-level themes describing the competencies reported in the literature for the effective and responsible use of LLMs in scientific research, which are available in the repository at \href{https://doi.org/10.5281/zenodo.21313656}{zenodo.21313656}. All coding decisions, code revisions, theme definitions, and final interpretations were performed and validated by the researchers. 

In addition to the manual analysis, a complementary thematic analysis was conducted with an LLM-based approach, whose implementation, including the model, version, parameters, and prompts, is documented in the artifacts (Section~\ref{sec:artifacts}). Operating independently over the 17 articles it had selected, the software tool extracted 199 competency occurrences and generated 9 themes, as summarized in Figure~\ref{fig:analysis_pipeline}(b). The AI-assisted analysis served as a complementary coding lens rather than an autonomous source of conclusions. Its themes were compared against the codes defined manually by the researchers, who reviewed and validated them and retained full responsibility for every coding decision, theme definition, and final interpretation.

Finally, by comparing the themes generated through the AI-supported analysis with those defined manually by the researchers, conceptual similarities and overlaps between categories were identified. The researchers refined the nomenclature produced by both analyses, grouping semantically related themes under a single label. This procedure reduced redundancies, standardized the terminology, and strengthened the coherence of the final thematic structure, consolidating the results into 8 final themes, which are presented in the Results.

The merging process was conducted by the two senior researchers, who independently reviewed the themes produced by both analyses and proposed equivalences based on conceptual overlap in definition and scope. Cases of disagreement were resolved through discussion until consensus was reached. No formal inter-rater agreement metric was calculated for this phase, as the decisions involved interpretive judgment about theme boundaries rather than binary classification.

The criterion for consolidation was conceptual equivalence. Themes from the manual and AI-assisted analyses that described the same competency were merged under a single label, while themes that appeared in only one of the analyses were kept as separate competencies. Applying this criterion to the seven manual themes and the nine AI-assisted themes resulted in eight final competencies. The complete mapping between the original themes and the eight consolidated competencies is available in the repository at \href{https://doi.org/10.5281/zenodo.21313656}{zenodo.21313656}.


\subsection{Threats to Validity}
Our findings should be interpreted considering the limitations inherent to the adopted method. First, as a rapid review, this study relied on Elicit and Google Scholar rather than curated bibliographic databases such as Scopus, Web of Science, the ACM Digital Library, and IEEE Xplore. Excluding these sources may have reduced both coverage and reproducibility, since peer-reviewed studies indexed only in them could have been omitted, and the semantic ranking performed by Elicit is not fully transparent, which further limits the replicability of the retrieval. Second, competency identification and coding involved qualitative interpretation, although independent screening, agreement assessment, and iterative discussions among the researchers were used to mitigate this threat. Third, the manual and AI-assisted analyses extracted competencies at different densities, approximately 3.7 instances per article in the manual analysis (147 across 40 articles) and 11.7 per article in the AI-assisted analysis (199 across 17 articles), and covered different numbers of studies. In addition, the pooled counts were not de-duplicated across the two analyses, so an instance identified in both contributed to each source. The pooled frequencies ($\Sigma n$) should therefore be read as an indication of the relative salience of each competency rather than a precise measure of prevalence or a count of unique instances. Finally, the identified competencies reflect a rapidly evolving body of literature and should be interpreted within the context of publications available between 2022 and 2025.

\section{Results}

Our analysis identified eight competencies associated with the use of AI in research. These competencies describe the knowledge, skills, responsibilities, and decision making processes reported in the literature as necessary for the effective, critical, and responsible use of AI throughout the research lifecycle. Table~\ref{tab:competencies} presents the identified competencies, their frequencies, definitions, and representative evidence extracted from the reviewed studies. Because the manual and AI-assisted analyses were pooled without de-duplication, the reported frequencies ($\Sigma n$) should be read as indicators of the relative salience of each competency rather than absolute counts of unique instances. The following sections describe each competency and discuss how they were characterized across the reviewed studies.

\subsection{Domain Expertise and Oversight of AI Outputs.} Domain expertise and oversight of AI outputs was the most frequently reported competency. This competency refers to the ability to critically assess AI generated outputs, verify information against external sources, identify inaccuracies and hallucinations, and retain responsibility for research decisions and conclusions. The reviewed studies suggest that this evaluation depends on sufficient domain expertise, since detecting subtle errors, superficial reasoning, or misleading claims requires subject-matter knowledge that may exceed what the AI tool itself provides. Researchers without an adequate command of the topic may accept plausible but flawed outputs, as the ability of the tools can exceed the skills of the person using them. Across the reviewed studies, domain knowledge was portrayed as essential for distinguishing meaningful contributions from superficial ones, recognizing which hypotheses merit further exploration, and arriving at ground-truth conclusions from AI responses. For example, a researcher using an LLM to summarize studies for a literature review would still need enough domain expertise to verify claims against the original sources, detect misrepresented findings, and judge whether the synthesis meaningfully advances the field before incorporating it into a manuscript.

\subsection{Metacognition and Decision Making Regarding AI Use}
The second most recurrent competency concerns the ability to determine when, where, and how AI should be used during research activities. This competency positions researchers as active decision makers who evaluate the appropriateness of AI support according to task characteristics, methodological requirements, and research goals. The literature frequently portrayed AI as a collaborator rather than a replacement for researchers. Studies highlighted the importance of maintaining humans in the loop, combining human and AI contributions, and deciding which tasks should remain under direct human control. Effective AI use therefore depends on researchers' ability to judge the suitability of AI support in different contexts. For example, a researcher might use AI to organize interview transcripts or generate preliminary themes while reserving theoretical interpretation and conclusion development for themselves.

\subsection{Ethics, Privacy, and Academic Integrity}
Ethics, privacy, and academic integrity emerged as another highly recurrent competency. This competency encompasses responsible AI use, protection of sensitive information, transparency regarding AI involvement, appropriate attribution of authorship, prevention of plagiarism, and awareness of legal and societal implications. The reviewed studies discussed ethical considerations from multiple perspectives, including data privacy, bias, ownership of AI generated content, responsible disclosure of AI use, and academic honesty. Collectively, these studies suggest that ethical considerations should accompany AI use throughout the entire research process rather than being addressed only at publication time. For instance, a researcher analyzing proprietary organizational data with AI tools would need to ensure that confidential information is protected, AI use is disclosed appropriately, and final interpretations remain under human responsibility.

\subsection{Prompt Engineering for Research}
Prompt engineering refers to the ability to formulate effective instructions that guide AI systems toward useful and reliable outputs. This competency includes defining objectives, providing context, specifying expected outputs, and iteratively refining interactions according to task requirements. Evidence related to this competency emphasized that prompt quality directly influences output quality. Studies described prompt engineering as a structured activity involving clear instructions, contextual information, iterative refinement, and explicit specification of desired outcomes. Researchers were frequently encouraged to refine prompts when initial outputs were unsatisfactory. For example, a researcher seeking themes in interview data may initially obtain generic results and subsequently refine the prompt by specifying the study context, participant population, and analytical objective.

\subsection{Reproducibility and Reporting of AI Use}
Reproducibility and reporting of AI use concerns the documentation and disclosure of how AI tools are employed during research activities. This competency includes reporting prompts, model versions, configurations, and AI assisted tasks to promote transparency, accountability, and reproducibility. The reviewed studies frequently emphasized the importance of documenting AI use. Reporting practices were discussed as mechanisms for enabling replication, supporting transparency, and helping readers understand how AI contributed to research outcomes. Several studies also referred to emerging institutional expectations regarding disclosure of AI use in scientific work. A research team using an LLM to assist with article screening in a literature review, for example, would be expected to document the prompts, model version, and validation procedures used throughout the process.

\subsection{Methodological and Experimental Design with AI}
Methodological and experimental design with AI refers to the ability to incorporate AI into research planning and study design. This competency includes defining research objectives, specifying constraints, describing datasets adequately, and designing workflows that account for the strengths and limitations of AI systems. The reviewed studies suggest that successful AI use depends substantially on the quality of the research design surrounding it. Evidence highlighted the importance of clearly formulated research goals, well defined datasets, structured workflows, and explicit methodological constraints. These elements influence the quality and reliability of AI generated outputs. For example, a researcher designing an AI assisted study would need to define research questions, describe the available data, and establish validation procedures before introducing AI into the workflow.

\subsection{AI Literacy and Technical Knowledge}
AI literacy and technical knowledge refers to understanding how AI systems function, including their capabilities, limitations, risks, and technical characteristics. This competency enables researchers to make informed decisions regarding tool selection, interpretation of outputs, and appropriate use of AI systems. The literature suggests that effective AI use requires more than operational familiarity with AI tools. Researchers were expected to understand model limitations, recognize potential risks, evaluate available tools, and determine whether specific systems were suitable for particular research tasks. Several studies also emphasized the need for formal training and AI related competencies within research education. For example, a researcher selecting an AI tool for qualitative coding would need to understand how the model operates, its limitations, and the potential implications of its outputs for the validity of the study.

\subsection{Data Analysis and Interpretation with AI}
The final competency concerns the use of AI to support quantitative, qualitative, and exploratory data analysis activities. It includes generating analytical code, supporting statistical analyses, identifying patterns, processing research materials, and assisting with interpretation while preserving human responsibility for final conclusions. Although less frequently reported than other competencies, several studies described AI as a useful tool for supporting analytical activities. Reported applications included generating code, supporting statistical procedures, identifying patterns in datasets, and assisting with exploratory analyses. At the same time, studies consistently emphasized that human researchers remain responsible for evaluating analytical quality and interpreting results. For example, a researcher using AI generated code to analyze software repository data would still need to verify the correctness of the analysis, assess the suitability of the analytical approach, and determine whether the resulting findings are supported by the evidence.

To improve traceability, Table~\ref{tab:competencies} is complemented by a competency-to-study matrix that maps each of the eight consolidated competencies to the primary studies from which its instances were extracted; the full matrix and the study identifiers are available in the artifacts (Section~\ref{sec:artifacts}).

\begin{table*}[!t]
\scriptsize
\caption{Competencies for AI-Assisted Research}
\label{tab:competencies}
\begin{tabularx}{\textwidth}{p{3.3cm}cp{4.3cm}X}
\hline
\textbf{Competency} & \textbf{$\Sigma n$} & \textbf{Description} & \textbf{Representative Evidence} \\
\hline
Domain Expertise and Oversight of AI Outputs & 123 &
Ability to critically assess AI-generated outputs, verify information, and identify inaccuracies, supported by sufficient domain expertise to recognize errors and retain responsibility for research conclusions. &
``Researchers should always evaluate critically the generated results and compare them with existing knowledge or expert opinions to ensure the validity of conclusions'' (ID41). ``Detecting such misleading statements can be challenging, especially when the abilities of the tools exceed the skills of the person using them'' (ID56). \\
\hline
Metacognition and Decision Making Regarding AI Use & 55 &
Ability to determine when, where, and how AI should be used during research activities and when human expertise remains necessary. &
``Its scientist-in-the-loop design prioritizes cognitive collaboration over full automation'' (ID151). ``We envision these technologies as augmenting rather than replacing human researchers'' (ID67). \\
\hline
Ethics, Privacy, and Academic Integrity & 53 &
Ability to use AI responsibly while considering privacy, authorship, transparency, plagiarism, and broader ethical implications. &
``Ethical concerns are well-documented in existing studies that highlight issues such as data privacy, misinformation, and unintended biases'' (ID49). ``Such initiatives can ensure the benefits of AI augmentation without compromising academic integrity'' (ID40). \\
\hline
Prompt Engineering for Research & 38 &
Ability to formulate, refine, and contextualize prompts to obtain reliable and relevant AI outputs. &
``Prompt engineering is the process of creating clear, concise and easily understandable prompts'' (ID12). ``GPT-4's performance can be improved by asking questions in an iterative manner or providing additional context'' (ID41). \\
\hline
Reproducibility and Reporting of AI Use & 29 &
Ability to document and disclose AI use, including prompts, tools, and configurations, to support transparency and reproducibility. &
``Authors are advised to retain the dialogue records to facilitate reviewers, editors and readers in replicating and understanding the process'' (ID12). ``The use of AI to generate texts needs to be explicitly acknowledged'' (ID63). \\
\hline
Methodological and Experimental Design with AI & 20 &
Ability to integrate AI into research design through clear objectives, constraints, workflows, and validation procedures. &
``Research goal: Scientist describes a research goal along with preferences, experiment constraints, and other attributes'' (ID142). ``The process reliability depends on the formulation of the research goal and the description of the dataset'' (ID5). \\
\hline
AI Literacy and Technical Knowledge & 16 &
Understanding of AI capabilities, limitations, risks, models, and technical characteristics required for informed use. &
``Researchers should be equipped with proper training to utilize chatbots and other AI tools effectively and ethically'' (ID40). ``The selection of a suitable base LLM model for finetuning is of utmost importance'' (ID93). \\
\hline
Data Analysis and Interpretation with AI & 7 &
Ability to employ AI to support analytical activities while preserving human responsibility for interpretation and conclusions. &
``Researchers consider using GenAI to support statistical analysis and help recognize patterns in data as good research practices'' (ID13). ``For data analysis and visualization tasks, we also report on the quality of the analysis and visualization'' (ID56). \\
\hline
\end{tabularx}
\end{table*}

\subsection{Answer to the Research Question}

Our findings indicate that the effective, critical, and responsible use of LLMs in scientific research requires competencies spanning the entire AI-assisted process, from understanding AI capabilities and limitations, designing workflows, and formulating prompts, to evaluating outputs, conducting AI-assisted analyses, ensuring transparency, and maintaining ethical responsibility. The eight competencies suggest that AI-assisted research is not simply about adopting new tools, but about developing expertise that combines technical knowledge, methodological rigor, critical thinking, and scientific judgment. Preparing researchers for this therefore represents an educational challenge requiring explicit training and guidance. In software engineering education, these findings point to the need to complement traditional research and technical competencies with AI literacy, human oversight, ethical decision making, transparency, and the critical evaluation of AI-generated outputs, so that future researchers and practitioners can use AI in ways that preserve research quality, integrity, and trustworthiness.

\section{Discussion}
\label{sec:discussion}

Although LLMs introduce new forms of assistance throughout the research lifecycle, our findings suggest that they do not alter the epistemic foundations of research. Rather than transferring responsibility for knowledge production to AI systems, the identified competencies indicate that researchers remain responsible for evaluating evidence, assessing validity, interpreting results, and justifying conclusions. AI-assisted research thus appears less as a replacement of traditional research expertise and more as a redistribution of where and how that expertise is exercised. The prominence of competencies related to domain expertise, oversight, metacognitive decision making, ethics, and reproducibility suggests that the challenge is not obtaining outputs from AI systems, but governing, evaluating, and appropriately integrating them into scientific work.

\subsection{Comparing Results with the Literature}
\label{sec:discussioncomparing}

Our findings align with existing discussions regarding AI-assisted research, which frequently emphasize concerns related to transparency, reproducibility, hallucinations, bias, source verification, scientific integrity, and researcher accountability \cite{meyer2023chatgpt, rahman2023chatgpt, lissack2024navigating, blau2024protecting, baltes2025guidelines}. Similarly, our results are consistent with traditional conceptions of research competence that emphasize critical evaluation of evidence, methodological rigor, ethical conduct, and scientific judgment as essential components of trustworthy research \cite{kitchenham2004evidence, kitchenham2009systematic, wohlin2007empirical, meerah2012measuring, reyes2021research}. The predominance of competencies related to domain expertise, oversight, ethics, and decision making suggests that these foundational research competencies remain relevant even as AI becomes increasingly integrated into research workflows.

At the same time, our findings differ from portions of the AI literacy literature that place stronger emphasis on understanding AI technologies and their technical operation \cite{ng2021conceptualizing}. Although AI literacy, technical knowledge, and prompt engineering emerged as relevant competencies, they were less recurrent than competencies associated with domain expertise, oversight, accountability, and methodological decision making. This pattern suggests that effective AI-assisted research may depend less on technical mastery of AI systems and more on researchers' ability to exercise judgment regarding when AI should be used, how outputs should be validated, and under what conditions generated content can be incorporated into scientific knowledge claims.

The novelty of this study lies in integrating discussions that frequently appear dispersed across different streams of literature. Previous work has addressed AI literacy, research integrity, responsible AI, prompt engineering, transparency, reproducibility, and methodological rigor largely as separate concerns. Our findings suggest that these topics collectively describe a broader competency profile required for AI-assisted research. By consolidating them into a single evidence-based framework, this study provides an integrated understanding of the knowledge, skills, responsibilities, and decision making processes needed to conduct research effectively, critically, and responsibly in AI-assisted environments.

\subsection{Implications}

\textbf{Research.} This study contributes an evidence-based consolidation of competencies associated with AI-assisted research. Existing literature has discussed opportunities, risks, applications, and guidelines for LLM use, but limited attention has been given to systematically identifying the knowledge, skills, responsibilities, and decision making processes researchers require to engage with these technologies responsibly. The identified competencies provide a foundation for future work on how they are developed, applied, and manifested across research contexts, disciplines, and methodological approaches, and may support competency models, researcher assessment approaches, and guidelines for responsible practice. More broadly, they contribute to discussions on the role of human judgment, accountability, and expertise in increasingly AI-mediated knowledge production.

\noindent \textbf{Software Engineering Education.} Preparing future researchers and graduate students for AI-assisted environments requires more than technical instruction on AI tools. By clarifying the knowledge, skills, responsibilities, and decision making processes associated with AI-assisted research, these findings may support instructors in designing research methods courses, graduate seminars, and AI literacy initiatives, and in defining learning objectives, structuring training activities, and setting expectations for responsible AI use across the research lifecycle. For students, particularly at the graduate level, they offer a clearer understanding of what is expected in increasingly AI-assisted environments. Incorporating these competencies into research training may help prepare researchers who engage with AI while maintaining the rigor, integrity, and accountability expected of scientific research.

\section{Conclusion}

This study investigated the competencies required for the effective, critical, and responsible use of Large Language Models throughout the scientific research process. Through a rapid review of studies published between 2022 and 2025, we identified eight competencies associated with AI-assisted research, spanning domain expertise and oversight, decision making regarding AI use, ethics and academic integrity, prompt engineering, reproducibility, methodological design, AI literacy, and data analysis. Our findings suggest that effective use of LLMs in research depends not only on understanding AI systems but also on maintaining the forms of judgment, accountability, and methodological rigor traditionally associated with scientific research. The prominence of competencies related to domain expertise, oversight, ethics, and decision making indicates that researchers remain responsible for assessing evidence, validating outputs, and justifying conclusions even when AI systems participate throughout the research lifecycle. Beyond identifying individual competencies, this study consolidates discussions that frequently appear dispersed across the literature on AI literacy, responsible AI, research integrity, reproducibility, and AI-assisted research practices. The resulting synthesis provides an evidence-based view of the knowledge, skills, responsibilities, and decision making processes associated with conducting research in increasingly AI-assisted environments. Findings also suggest that preparing researchers for these environments requires attention not only to technical aspects of AI use but also to the development of competencies that support critical engagement with AI-generated outputs and responsible integration of AI into research workflows. \textbf{Future Work.} Future studies may investigate how these competencies manifest across different disciplines, research methodologies, and stages of the research lifecycle. Additional research may also explore how researchers develop these competencies over time and how they can be assessed and supported through training initiatives and institutional guidance. Such investigations may contribute to the development of competency models, researcher preparation strategies, and guidelines that support the responsible use of AI in scientific research.

\section{Artifacts}
\label{sec:artifacts}

All materials underlying this review are available in the replication package at \href{https://doi.org/10.5281/zenodo.21313656}{zenodo.21313656}, including the complete search strings for Elicit and Google Scholar, the screened studies, the coding files of both analyses, a competency-to-study traceability matrix, and the full specification of the AI-assisted procedures (model, version, parameters, and prompts).

\section*{Declaration on the Use of Artificial Intelligence}

Generative AI tools supported parts of the qualitative data analysis, as described in the methodology, and assisted with grammar and language revision of the manuscript. All AI-generated outputs and suggestions were reviewed, verified, and edited by the authors, who retain full responsibility for all analytical decisions, findings, and conclusions.

\bibliographystyle{sbc}
\bibliography{sbc-template}

@article{lissack2024navigating,
  title={Navigating the future of large language models in scientific research: Opportunities, challenges, and ethical considerations},
  author={Lissack, Michael and Meagher, Brenden},
  journal={Challenges, and Ethical Considerations (September 02, 2024)},
  year={2024}
}

@inproceedings{reddy2025towards,
  title={Towards scientific discovery with generative ai: Progress, opportunities, and challenges},
  author={Reddy, Chandan K and Shojaee, Parshin},
  booktitle={Proceedings of the AAAI conference on artificial intelligence},
  volume={39},
  number={27},
  pages={28601--28609},
  year={2025}
}

@misc{blau2024protecting,
  title={Protecting scientific integrity in an age of generative AI},
  author={Blau, Wolfgang and Cerf, Vinton G and Enriquez, Juan and Francisco, Joseph S and Gasser, Urs and Gray, Mary L and Greaves, Mark and Grosz, Barbara J and Jamieson, Kathleen Hall and Haug, Gerald H and others},
  journal={Proceedings of the National Academy of Sciences},
  volume={121},
  number={22},
  pages={e2407886121},
  year={2024},
  publisher={National Academy of Sciences}
}

@article{baltes2025guidelines,
  title={Guidelines for empirical studies in software engineering involving large language models},
  author={Baltes, Sebastian and Angermeir, Florian and Arora, Chetan and Bar{\'o}n, Marvin Mu{\~n}oz and Chen, Chunyang and B{\"o}hme, Lukas and Calefato, Fabio and Ernst, Neil and Falessi, Davide and Fitzgerald, Brian and others},
  journal={arXiv preprint arXiv:2508.15503},
  year={2025}
}

@article{rahman2023chatgpt,
  title={ChatGPT and academic research: A review and recommendations based on practical examples},
  author={Rahman, Mizanur and Terano, Harold Jan R and Rahman, Nafizur and Salamzadeh, Aidin and Rahaman, Saidur},
  journal={Journal of Education, Management and Development Studies},
  volume={3},
  number={1},
  pages={1--12},
  year={2023}
}

@article{meyer2023chatgpt,
  title={ChatGPT and large language models in academia: opportunities and challenges},
  author={Meyer, Jesse G and Urbanowicz, Ryan J and Martin, Patrick CN and O’Connor, Karen and Li, Ruowang and Peng, Pei-Chen and Bright, Tiffani J and Tatonetti, Nicholas and Won, Kyoung Jae and Gonzalez-Hernandez, Graciela and others},
  journal={BioData mining},
  volume={16},
  number={1},
  pages={20},
  year={2023},
  publisher={Springer}
}

@article{chen2020artificial,
  title={Artificial intelligence in education: A review},
  author={Chen, Lijia and Chen, Pingping and Lin, Zhijian},
  journal={IEEE access},
  volume={8},
  pages={75264--75278},
  year={2020},
  publisher={Ieee}
}

@inproceedings{kirova2024software,
  title={Software engineering education must adapt and evolve for an llm environment},
  author={Kirova, Vassilka D and Ku, Cyril S and Laracy, Joseph R and Marlowe, Thomas J},
  booktitle={Proceedings of the 55th ACM technical symposium on computer science education v. 1},
  pages={666--672},
  year={2024}
}

@article{santos2026llm,
  title={LLM Use, Cheating, and Academic Integrity in Software Engineering Education},
  author={Santos, Ronnie de Souza and Santos, Italo and Bento, Mariana and Destefanis, Giuseppe and Magalh{\~a}es, Cleyton and Wessel, Mairieli},
  journal={International Conference on the Foundations of Software Engineering (FSE) },
  year={2026}
}

@article{ng2021conceptualizing,
  title={Conceptualizing AI literacy: An exploratory review},
  author={Ng, Davy Tsz Kit and Leung, Jac Ka Lok and Chu, Samuel Kai Wah and Qiao, Maggie Shen},
  journal={Computers and Education: Artificial Intelligence},
  volume={2},
  pages={100041},
  year={2021},
  publisher={Elsevier}
}

@article{reyes2021research,
  title={Research competencies mediated by technologies: A systematic mapping of the literature},
  author={Reyes, Carlos Enrique George and Morales, Leonardo David Glasserman},
  journal={Education in the Knowledge Society (EKS)},
  volume={22},
  pages={e23897--e23897},
  year={2021}
}

@article{meerah2012measuring,
  title={Measuring graduate students research skills},
  author={Meerah, T Subahan Mohd and Osman, Kamisah and Zakaria, Effendi and Ikhsan, Zanaton Haji and Krish, Pramela and Lian, Denish Koh Choo and Mahmod, Diyana},
  journal={Procedia-Social and Behavioral Sciences},
  volume={60},
  pages={626--629},
  year={2012},
  publisher={Elsevier}
}

@inproceedings{wohlin2007empirical,
  title={Empirical software engineering: Teaching methods and conducting studies},
  author={Wohlin, Claes},
  booktitle={Empirical Software Engineering Issues. Critical Assessment and Future Directions: International Workshop, Dagstuhl Castle, Germany, June 26-30, 2006. Revised Papers},
  pages={135--142},
  year={2007},
  organization={Springer}
}

@article{pizard2021training,
  title={Training students in evidence-based software engineering and systematic reviews: a systematic review and empirical study},
  author={Pizard, Sebasti{\'a}n and Acerenza, Fernando and Otegui, Ximena and Moreno, Silvana and Vallespir, Diego and Kitchenham, Barbara},
  journal={Empirical Software Engineering},
  volume={26},
  number={3},
  pages={50},
  year={2021},
  publisher={Springer}
}

@article{kitchenham2009systematic,
  title={Systematic literature reviews in software engineering--a systematic literature review},
  author={Kitchenham, Barbara and Brereton, O Pearl and Budgen, David and Turner, Mark and Bailey, John and Linkman, Stephen},
  journal={Information and software technology},
  volume={51},
  number={1},
  pages={7--15},
  year={2009},
  publisher={Elsevier}
}

@inproceedings{kitchenham2004evidence,
  title={Evidence-based software engineering},
  author={Kitchenham, Barbara A and Dyba, Tore and Jorgensen, Magne},
  booktitle={Proceedings. 26th International Conference on Software Engineering},
  pages={273--281},
  year={2004},
  organization={IEEE}
}

@article{trinkenreich2026taking,
  title={Taking a Pulse on How Generative AI is Reshaping the Software Engineering Research Landscape},
  author={Trinkenreich, Bianca and Calefato, Fabio and Blincoe, Kelly and Wivestad, Viggo Tellefsen and Alves, Antonio Pedro Santos and Ara{\'u}jo, J{\'u}lia Cond{\'e} and Ara{\'u}jo, Marina Cond{\'e} and Tell, Paolo and Kalinowski, Marcos and Zimmermann, Thomas and others},
  journal={arXiv preprint arXiv:2604.11184},
  year={2026}
}

@article{kozov2024practical,
  title={Practical Application of AI and Large Language Models in Software Engineering Education.},
  author={Kozov, Vasil and Ivanova, Galina and Atanasova, Desislava},
  journal={International Journal of Advanced Computer Science \& Applications},
  volume={15},
  number={1},
  year={2024}
}

@inproceedings{trinkenreich2025get,
  title={Get on the train or be left on the station: Using llms for software engineering research},
  author={Trinkenreich, Bianca and Calefato, Fabio and Hanssen, Geir and Blincoe, Kelly and Kalinowski, Marcos and Pezz{\`e}, Mauro and Tell, Paolo and Storey, Margaret-Anne},
  booktitle={Proceedings of the 33rd ACM International Conference on the Foundations of Software Engineering},
  pages={1503--1507},
  year={2025}
}

@incollection{cartaxo2020rapid,
  title={Rapid reviews in software engineering},
  author={Cartaxo, Bruno and Pinto, Gustavo and Soares, Sergio},
  booktitle={Contemporary empirical methods in software engineering},
  pages={357--384},
  year={2020},
  publisher={Springer}
}

@inproceedings{cruzes2011recommended,
  title={Recommended steps for thematic synthesis in software engineering},
  author={Cruzes, Daniela S and Dyba, Tore},
  booktitle={2011 international symposium on empirical software engineering and measurement},
  pages={275--284},
  year={2011},
  organization={IEEE}
}

@article{gwet2008computing,
  title={Computing inter-rater reliability and its variance in the presence of high agreement},
  author={Gwet, Kilem Li},
  journal={British Journal of Mathematical and Statistical Psychology},
  volume={61},
  number={1},
  pages={29--48},
  year={2008},
  publisher={Wiley Online Library}
}

@article{cohen1968weighted,
  title={Weighted kappa: Nominal scale agreement provision for scaled disagreement or partial credit.},
  author={Cohen, Jacob},
  journal={Psychological bulletin},
  volume={70},
  number={4},
  pages={213},
  year={1968},
  publisher={American Psychological Association}
}

@book{krippendorff2018content,
  title={Content analysis: An introduction to its methodology},
  author={Krippendorff, Klaus},
  year={2018},
  publisher={Sage publications}
}

@article{landis1977measurement,
  title={The measurement of observer agreement for categorical data},
  author={Landis JRKoch, GG},
  journal={Biometrics},
  volume={33},
  number={1},
  pages={159174},
  year={1977}
}

\end{document}